   \documentstyle[aps]{revtex}
 \include{psfig}
\begin{document} 
\title{
The lid method for exhaustive exploration of
metastable states of complex systems 
} 
\author{
Paolo Sibani\\
 }
\address{
 Fysisk Institut, Odense Universitet\\
Campusvej 55,\ DK5230 Odense M \\
Denmark
}
\author{
Ruud van der Pas\\
}
\address{ European High-Performance Computing Team, Silicon Graphics\\
 Veldzigt 2A, 3454 PW De Meern \\
The Netherlands} 
\author{
J. Christian Sch\"{o}n\\
 }
\address{
 Institut f\"{u}r Anorganische Chemie, Universit\"{a}t Bonn\\
Gerhard-Domagk-Str. 1,\ D-53121 Bonn\\
 Germany 
}

\date{\today} 
\maketitle
%\vspace{-0.5cm}

\begin{center}
 {\bf Abstract} 
 \end{center}
 
\begin{footnotesize} 
The `lid' algorithm  is  designed 
for the  exhaustive exploration  of
neighborhoods of local energy minima
 of   energy  landscapes.
This paper describes an  implementation
of the algorithm, including issues of parallel
  performance and scalability. To 
illustrate the versatility of the approach  and to  
stress   the  common features  present in landscapes of 
quite different systems,    we present selected  results 
for   1) a  spin glass, 2) a ferromagnet,  
3)  a covalent network model for   glassy systems, and  
4) a polymer. 
 The exponential nature of the local density of states
found in these systems  and its relation to the 
ordering transition is briefly commented upon. 
 \end{footnotesize}
 
\section{Introduction} \label{introduction}

The state space $\bf{S}$ of a complex systems  
  together with a scalar function
$E: \bf{S} \rightarrow \bf{R}$, is often  
described as a `landscape'\cite{Landscape97}.   
 For discrete state spaces, a landscape is  a
 graph, whose  nodes represent
 the states and  where edges   connect   neighbor states. 
 Each node is characterized by a scalar function, $E$, which
 in physical systems is usually the  energy of a state,
 while  in other cases it can be
 a  cost or a fitness.  Often the system dynamics  
 takes the form  of a Markov process with   states in  the  landscape 
 and transitions among neighbors. In  thermally activated  processes
  hopping    to states of higher energy is a very rare event at low
 temperatures.  In this regime  
 the system remains trapped  for very long times 
  within relatively small sets of states    
 surrounding  local energy minima.  
The  relation between low temperature relaxation 
dynamics and the  geometrical properties
 of  the traps, which can be studied by the 
lid algorithm presented
in this paper, is   a topic  of great current 
interest. Several theories for the behavior   
of complex systems at low temperature, e.g. aging,  have been
advanced\cite{Hoffmann88,Sibani89,Bouchaud92,Kurchan96,Sibani97}, which 
build on  assumptions about   state space geometry. In 
chemistry, low-energy configurations of 
clusters\cite{Berry93,Berry94,Kunz94,Wales96}, 
proteins\cite{Dobson98,Onuchic97}, polymers\cite{Becker97} 
and solids\cite{Schon96c,Putz98,Schon97}
may represent metastable compounds, whose  
reaction pathways and lifetimes are of 
considerable importance.
Finally, heuristic   optimization techniques
based on annealing \cite{Mobius97}
 may benefit from
any insight on generic properties of the low energy 
part of the landscapes \cite{Schon97b}.

The  `lid' algorithm  provides exact  geometrical information by
 exhaustively   visiting subsets of state space, which are    
 characterized by two quantities: 
a low energy `reference state', $\psi$ , 
  and a `lid', $L$.
The enumeration starts with   $\psi$,   
and covers all those states which are connected 
to $\psi$  by paths   never exceeding  the energy level  $L$.
Clearly, the set of states registered,  ${\cal P}_{\psi,L}$,
 henceforth 
called a `pocket', is likely to behave as  a dynamical trap 
in a thermal relaxation process.  Roughly, one expects the 
trajectory to remain confined to the pocket for   time scales 
of order $\exp(L/T)$, where $T$ is the temperature 
and  the Boltzmann constant is set to one (the escape time
 may exceed this value considerably, if large entropic barriers
  are present\cite{Schon97}). 
If a deep pocket is metastable, it is 
possible to study the relaxation behavior in its interior
(independently of the 
rest of the energy landscape), either at  the microscopic  level 
 \cite{Sibani93} or using a coarse-grained "lumped" model of the 
pocket\cite{Schon96}.
 
Following a discussion  of the  algorithm, 
 we   present   selected results  from 
four rather different  applications. This  
 demonstrates the versatility of the method  and
 stresses  the interesting - and not  
 widely recognized fact -  
 that important features of landscapes are  common 
 to  rather different   systems.
 A   more detailed discussion of some of the examples 
 can be found in separate publications\cite{Schon97,Sibani98,Schon98}.
 Similarly, we refer to the 
literature for implementations and applications of the lid method 
to  continuous (rather than discrete)
landscapes\cite{Schon96,Schon96b}. A different approach to 
the exploration of complex landscape based on 
 branch and bound methods can be found in ref.~\cite{Klotz98}.

\section{The lid method}  \label{method}
  
 To visualize how the lid method works, imagine  
 water welling up at a local minimum $\psi$ of the landscape.
 For concreteness we take the energy of this state, $E(\psi)$,
  to  be equal to zero. The height of the water level above
$\psi$ then  corresponds  to the value of the lid $L$. 
Initially, $\psi$ will be the deepest point of a
 lake, but, eventually,  another point $\psi'$ with lower energy 
 \mbox{$E(\psi')< E(\psi)$}   
 might  become  submerged.  If this happens,  
  a watershed is crossed and  the 
 pocket containing  $\psi'$ is flooded. 
 The smallest lid value at which the overflowing occurs  
defines  the depth of the pocket  centered at $\psi$. 
 
 The amount of landscape submerged, or volume,   
  ${\cal V}_{\cal P}(L)$ provides   
 a   simple measure of the shape of the pocket
 as a function of the lid.  Further information is provided 
 by the height distribution of the submerged points,
  i.e., the local density of states ${\cal D}_{\cal P}(E;L)$, by  the number
of accessible minima,  ${\cal M}_{\cal P}(L)$, and by their
distribution ${\cal DM}_{\cal P}(E;L)$. 
 Depending on the problem at hand,   additional   questions can be asked:
 what is the height of the lowest saddle which must be reached before the water 
 can flow to the `sea' - i.e.  - before the  flooded part of state space
 percolates.  And how does this height scale with some important 
 parameter of the  problem - e.g. its size (number of atoms or spins) - or 
 the value of an external field. All these properties are local, since
  they are independent of other properties of the landscape 
   outside the pocket  itself.

 \section{The search algorithm} \label{algorithm}
The application of the lid algorithm to 
a given problem involves task-dependent as well as
task-independent  procedures. To the former class belong  
  the evaluation of the energy function,    the implementation
  of the move class   and   the
  coding of the  configurations.  
   The task independent features are
  the  generation and storage of the configurations, and their
  subsequent retrieval from a suitably organized data base.
    The link  between the two types of tasks is 
     the coding of the system configurations 
 as   binary strings. (Note that e.g. an  Ising
model is already naturally coded this way).

 The purpose of the  search  is to enumerate all those states of
  the system which can be reached by repeated applications of   elementary 
  moves,  starting at   $\psi$  and  without ever exceeding a preset value, 
  $L$,  of the energy.
 To understand  how the enumeration works, one may 
  identify   three  disjoint sets of states.
   The first set, {\bf A},  consists of 
states which have been visited, 
and whose neighbors  have all been 
visited. 
 The second set, {\bf B}, includes points which have been visited, but where
 not all the neighbors have been seen. 
 Finally, the third set, {\bf C}, includes all
 the accessible but as yet undiscovered  points. Initially, 
 {\bf B} contains just one point - the reference state $\psi$, {\bf A}
  is empty, and 
 {\bf C} contains a finite but unknown  
  number of points.  The program terminates 
  when  {\bf B} and {\bf C} 
  are empty, or earlier, if a lower lying reference state
   is discovered during the search.  
  
 In principle the enumeration task can be accomplished using two data
 structures: 1) a data base  
 containing  the visited states, henceforth referred to as  a `tree'  
 and 2) a linked list containing pointers to the elements of the data base.
 This list is later also referred to as a `buffer'.  
 The data base is ordered according to the  value
 of a {\em tag}, which is the value in base ten 
 of the binary string encoding each configuration. 
 The tree and the buffer
 will be collectively called a `search structure'.
 
 The progress  of the algorithm is given by a position marker
   along the list: states to the left of the marker are 
 of class {\bf A}, those to the right are of class  {\bf B}.
  The subroutine, 
 {\em generate}, takes as its  `current'   state 
 the first available state of class  {\bf B}  and calculates  
  all  its neighbors. This is done in several
steps: First, the configuration is decoded, i.e.
the actual configuration is calculated from the   tag.
 Then the neighbors are created and their energy is computed.
 States above the lid are immediately discarded.
 If the system possesses translational 
 or other symmetries, each of the remaining states  
 must first be  brought into  
 a unique representative configuration. When this is done,
 the  configuration is   encoded into a binary string,  
whose  tag  is computed  and then used to check 
the new configuration against    states already found.
When appropriate,
the configuration  is   appended to the data base and 
a corresponding pointer is added to the linked list, to the
immediate  right
 of the current marker  position, i.e. in class {\bf B}.
 Upon return from  {\em generate}, the 
 current state is updated,  by moving the pointer 
 one step to the right along the list.
 Initially, the number of  {\bf B} states greatly increases. As the calculation 
 progresses, fewer and fewer new states are found below the lid, and, 
 finally, the search  terminates when the current position of the marker 
 reaches  the end of the list.

 In pseudo-code, the conceptual, if not the actual,   structure of the 
(search part of the) algorithm is
 quite simple.
 \newpage  
\begin{small}
\begin{enumerate}

\item Read input data.
\item Initialize structures:
                \begin{enumerate}
                 \item Store reference configuration in binary search tree.
                \item Initialize the linked list {\bf buffer}. The first and only record
         in the buffer has two elements:    a pointer  to the reference configuration
         and a null pointer. The latter  will later point to the next element
          of the buffer.
   \item Initialize  a pointer, {\bf current}, to  point to the first position in 
        the buffer.  
  \end{enumerate} 
\item Generate neighbors to the current configuration.\\
        {\em If} {\bf  current}  is not null:  
  \begin{enumerate}
      \item Create a list of all neighbor states  of current configuration.
       Assign to each of these  a unique {\bf tag}.
      \item Remove from the list those neighbors which have energy above the lid.
      \item Remove from the list those states  which have the same tag as 
            states previously stored in the binary tree.
  \end{enumerate}
      {\em else} Exit the program.  
\item Append to the binary tree and update the buffer:
 \begin{enumerate}
      \item Append each new state to the binary tree,   ordered according 
            to the value of {\bf tag}. 
      \item For each append, insert a new record in the buffer, right after
            the `current' position. The two  elements of the record 
            are a pointer to the configuration just appended, and a pointer to the
            next record in buffer.     
  \end{enumerate}

\item Move the {\bf current} pointer one step forward along the buffer.

\item Go to point 3.
\end{enumerate}
\end{small} 
In the actual implementation of the code
it is important to minimize the
number  of {\em malloc} calls and  of  cache misses
by allocating memory in large blocks.
The  parallel implementation of the code described
in  section \ref{parallel} 
 is rather more complicated, but 
  provides a substantial speed-up.
 
\section{Parallel implementation} \label{parallel}
   
As the size of the data base which has to be
managed by the lid algorithm can be considerable
for large lids, i.e. up to  
 hundreds or even thousands 
of Megabytes of RAM, there is  
 a strong motivation 
to increase   the speed of execution   by
 parallelizing the search.
 A very coarse-grained parallelism,
e.g. the  simultaneous execution of multiple runs  with different input 
parameter values is not  a viable option due to the very large memory 
requirements. Two lower  levels of parallelization were considered:
1) Generate  the  neighbors of each configuration   in parallel, as 
     described in point 3 of the algorithm in section \ref{algorithm}.
2) Define  $N$   buffers and $N$ binary trees   
and perform the   search
 and append procedure in parallel.
  As the first   approach does not 
 lead to any  speed-up   we shall concentrate
  on the second, 
which   parallelizes quite  successfully.
 
\subsection{Parallel structure}
In order to balance  the search load among $NP$   processors, 
  each configuration is assigned a  positive integer $ID$,
  with    $ 0 \leq $ $ID$ $ < N  $.
Each $ID$ identifies a parallel thread, and each processor
runs $NP/N$   threads.  Each thread has its own  buffer and 
  search tree, which together constitute   
 a search structure.  
To obtain   a uniform distribution of the load across
the processors, the    values of 
the assigned $ID$'s   should be  uniformly distributed in the interval
$[ 0 , N ]$. Within the above  load-sharing scheme the scalability 
of the algorithm 
can be negatively affected by 1) the overhead for communication
along $N (N-1)$ different channels, and 2) deviations from perfect 
internal load balance. Our analysis shows that  the
main detrimental effect stems from 2), at least for moderate values of $NP$.

The  program initialization procedure is  similar to   
 the sequential case. All   the $N$ search structures 
 will  initially be
empty, except for the zero'th one,
 which must contain  the reference state.
The parallel  program includes
points 3) and 4) of the algorithm in section \ref{algorithm}
  as well as  a supplementary part dealing 
with the   communication among different threads.
The need of communicating arises because 
 the  $ID$ of  a configuration
 can  differ from those of its neighbors.
 Hence,  only a fraction
of the newly generated  states  
can  be   handled by the same thread as their parent. 
The rest is  temporarily stored in
 a square array of linked lists
$  mail $, where $ mail[i][j]$ contains states
generated by thread  $j$  which   have to  be handled
 by thread  $i$.

 All parallelism was  implemented at  a high level, using 
so called \#pragma constructs 
\cite{refComp}. These  are ignored on a single processor machine,
which makes the program  easily portable to different architectures. 
 The relevant parallel part of the code  looks as follows:  
\newpage 
\begin{small}
\begin{tabbing}
while   ( \= not\_all\_threads\_idle) \\
\{ \\
\>   \#pragma paral\=lel  (start  parallel \= region)   \\
\>   \{       \\
\> \>         \#pragma pfor (.....)\\
\> \>           for ( each thread i, i=0, ... N)  \\
\> \>         \{   \\
\> \> \>              {\em get\_from\_mail}  \=   (read mail from other threads)\\ 
\> \> \>              {\em s\_search} \>  (check  if the received states are \\
\> \> \>                           \>  already present and update if needed )\\ 
\> \> \>              while (  more configurations are in buffer i  ) \\
\> \> \>        \{  \\  
\> \> \> \>         for current \= configuration: \\     
\> \> \> \>           {\em generate} \>( generates all the neighbors of current \\
\> \> \> \> \>         configuration,  calling the following three procedures:\\       
\>\>\>\>{\em create} \> (creates neighbor configurations and assigns their ID's \\
\> \> \> \> \>                  for the spin glass case, this routine is called \\
\> \> \> \> \>                   {\em flip} )  \\ 
\> \> \> \>          {\em energy} \> (energy calculation - for spin glass called\\
\> \> \> \> \>                  {\em init\_energy3D} ) \\
\> \> \> \>                     {\em s\_search} \> (see above) )\\         
\> \> \>         \} \\
\> \>           \} \\
\> \>         {\em mp\_end\_pdo}\\
\> \>         \hspace{0.3cm} {\em mp\_barrier}\\
\> \>         \hspace{0.6cm} {\em  mp\_barrier\_nthreads}\\
\> \>        \#pragma one processor \\
\> \>         {check if all threads are done } \\
\> \>              {\em mp\_barrier} \\
\> \>              \hspace{0.3cm}   {\em mp\_barrier\_nthreads} \\
\>     \} \\
\}
\end{tabbing} 
\end{small}
 
In the above  fragment of pseudocode, all routines starting with "mp" are 
not user written. Calls to these routines from the
SGI thread library are inserted by the compiler 
  to steer the parallelization.
 The {\em \#pragma parallel} construct  causes the  `for-loop' 
 to be executed in parallel. After
completion of the `while-loop',  all threads are synchronized again.
 Then, one of the threads
 checks whether the algorithm is finished or not, and  sets a shared
flag accordingly. In this way, all other threads will be notified about
the current status, and either continue with their `while-loop' or terminate.
In addition to this explicit parallelism, a  lock  is used
to implement the earlier mentioned "mail" mechanism, which
enters in the routine 
 {\em get\_from\_mail}. 
Basically, each   thread  monitors a specific memory location, 
in order to check for any incoming
states. If there are such states, the thread will 
raise a flag (to stop new mail from flowing in), empty the mail
queue with the states into a memory location 
which it owns,   reset the flag and continue. 
 
\subsection{Parallel performance analysis}\label{performance}
 
To assess  the efficiency of the parallel implementation,  we   first ported 
the   program  to a Silicon 
Graphics Origin2000 system with MIPS R10000 processors 
running at 250MHz\cite{refO2000}. 
This machine has  a 64 bit  shared memory architecture i.e. the entire
 address space is 
accessible to all threads. 
We mainly used the spin-glass model for the tests, since it  has an easily
evaluated energy function and hence a lower  one-processor load and a 
higher communication load relative to the glass and polymer problem.
The parallel scalability of most other problems is expected to be higher
than the spin-glass case. 

In the spin glass   tests, the length of the parallel `for-loop' was $16$ and  the number
of processors was varied from $1$ to $16$. Two series of runs were
performed, with the  lid value  set
 to $0.038$   and to $0.046$ respectively.
The corresponding memory requirements were  
$200$ Mbytes and    $2$ Gbytes. 
 
Table~\ref{1}   lists the elapsed times in seconds  
  together with   two parallel speed-up metrics, for the 
test case with the smallest lid.  We
have used the uniprocessor elapsed time as a reference. Under column
"Rel. Speed-up" we list the speed-up obtained when doubling the number
of processors. Ideally this number should be $2$. Under column "Cum. Speed-up"
we define the speed-up on $P$ processors to be $T(1)/T(P)$. In the ideal case
this should be $P$. 
The last column contains the estimated performance using
 Amdahl's law\cite{Amdahl}. 
Briefly,  Amdahl's law assumes that the uni-processor elapsed
time $T(1)$   can be split in an optimizable
 and a non-optimizable
part, say:  $T(1) = f T(1) + (1-f) T(1)$, where  $f$ is the fraction of the
time that can benefit from optimization (which  in our case  is achieved 
through parallelism).
The elapsed time $T(P)$ on $ P$ processors is then
 $T(P) = (f/P) T(1) + (1-f) T(1)$.
By measuring  $T(1)$ and $T(2)$  the equation can be solved for $f$,
whence $T(P)$  can be (crudely) estimated for any  $P$.
In our case the procedure yields  $f = 0.81$. 

As can be seen, the simple Amdahl  model  severely underestimates
the performance,   probably due  to the 
anomalous behavior with $2$ processor  as 
 further discussed in the sequel. 
The parallel performance was  analyzed in more detail using the IRIX SpeedShop
profiling environment \cite{refSpeed}. Among other things, this tool gives the 
elapsed time for every function executed by each thread, thus clarifying  the 
scalability of different parts of the program. 
The results are presented in Table \ref{2}.
The function {\em s\_search} does not appear in this table because  it was
inlined by the compiler. 
Since  a run with a lid value of $0.038$ constitutes  a  rather small job,
 where the parallelization 
overhead can  be relatively dominant,  we   also ran the program 
for a lid value of $0.046$ requiring  slightly  more than $2$GB of main memory. 
The relative timings, plus associated metrics, are listed in table \ref{3}.
As   expected, we achieve a higher parallel efficiency for this problem
size. And, again, Amdahl's
law strongly underestimates the performance for
the large processor numbers. 
In table \ref{4} we list the profile information obtained on 1, 2 4, 8 and 16
processors.  The speed-up values for the user routines are given in table 
\ref{5}. These routines account for 99\% of the total computation time on 
a single processor. 
The corresponding speed-up values are given in table \ref{5}.
One observes that the user routines parallelize rather well albeit 
  not perfectly.
Also,  the routines  do not scale by the same factor, which of
course  limits the 
overall scalability of the program. 
As expected,  the function  {\em get\_from\_mail}, which 
implements the exchange of (a rather small amount 
of)  data among the $N $ threads is the 
poorest performer in terms of scalability, as  
the costs of locking the mailbox
must become  dominant for high $N$.  

From table \ref{4} we observe that the cost of the 
pragma generated barrier function  
decreases from its maximum at $N=2$. The SpeedShop profiler was also used to 
analyze which part of the program is actually responsible for the heavy usage 
of the barrier function. It was found to be in the parallelized for-loop i.e. 
the first barrier construct in the code fragment shown above is mainly 
responsible for the times given in table \ref{4}. The same type of behavior
appears more clearly in the   performance of the lid algorithm on
the  network glass model  described in 
section\ref{structural}. 
Unlike the spin-glass model, the energy evaluation and the
 coding-decoding of the string representation of a configuration are computationally 
demanding. Indeed,  the eight most time consuming routines 
perform these very  tasks. The elapsed times in seconds and the other metrics are 
listed in Table\ref{6}. For reasons that will be explained below, and in contrast 
with previous estimates, we have used $T(1)$ and $T(4)$ to estimate the fraction of 
parallelized code entering the Amdahl rule.
As is clear from the data, the performance on 2 processors sticks out in
a negative way, even though a profile of the user routines  (not included),
 shows that these
scale perfectly.  The fact that 
the cost of the barrier reaches a peak for $2$ 
processors  probably stems from an imbalance in the 
workload assigned by the algorithm to each processor:
As all processors are synchronized 
at the barrier located at the end of the parallelized {\em for} loop,
 the slowest processor sets the overall pace. 
Arguably, the  
 imbalance   increases with  the
 workload per processor and 
  the cost of synchronization is  highest when 
  the load per processor peaks. For fixed
 total load this  happens  at $NP=2$, as observed.  As a further check 
 we performed  tests for a series of increasing lids
 (i.e. increasing total workloads) and found a systematic
 deterioration of the parallel performance with two processors,
 but no negative effects with  eight processors.
Finally, we note that the second of the compiler inserted barriers,
which is located at the end of the one processor section of the code, 
has a minimal impact on scalability. This is because the processors,
having  just been synchronized by the previous barrier,
spend very little time at this part of the code.   
  
In summary,  we see that: 1) for a spin-glass problem with $16$ threads
  the shortest turnaround time   is obtained on  $16$ processors,
but the best parallel efficiency on $8$ processors. 2) Due 
to some non-scalable parts 
in the program, a slight load imbalance
  within the algorithm and the cost of synchronization itself, the scalability
  of the program is not perfect. However, on a medium sized production problem,
 the current implementation 
  still achieves a speed-up of up to $9.3$ on $16$ processors. 
 Scalability is slightly better on the more 
 computation intensive glass problem. In general, 
we  expect  that on larger problems and on problems with
a higher one-processor load,  the parallel efficiency will 
 improve further. As the cost of synchronization seems
 quite substantial, high latency networks (typically a 
cluster of machines) would probably not handle   this type 
 of problems efficiently.
A more asynchronous implementation of the algorithm is likely 
to reduce idle time and further enhance scalability. The SGI parallel   
environment does provide means to implement this, but the possibility
has not yet been investigated by the authors. 
 
\section{Applications} \label{applications}
To illustrate the versatility of the lid algorithm,
and highlight the presence of common features across
different systems, we briefly  discuss four    applications.
A   detailed discussion of the physics
is given in Ref.~\cite{Sibani98} for the spin glass case,
in Ref.~\cite{Schon98} for the 2d-network case, and
finally in Ref.\cite{Schon97} for the polymer case. 
Here we just stress the observation that   
the scale parameter of the exponential  growth of 
e.g. the local density of states  
 can be identified with the  temperature where the trap 
looses its thermal metastability. 
In some cases, (e.g. the spin-glass
and the ferromagnetic models)  
this temperature turns out  
to be quite close to the  actual 
 critical temperature of the system.

\subsection{Spin glass} \label{spin}
A set   of $\cal N$ Ising spins,
$\sigma_i = \pm 1$, is
placed on a  3D cubic lattice with periodic
boundary conditions.\ 
 The energy of the $x$'th configuration,
$0< x \leq 2^{\cal N}$,     
 is defined by the well known Hamiltonian\cite{Edwards75} 
\begin{equation}
E(x)= - \frac{1}{2} \sum_{i,j} J_{ij} \sigma _i(x) \sigma _j(x),
\label{Ham}
\end{equation}
where  $J_{ij}= J_{ji}$  and where $J_{ij}\neq 0$
only if spins  $i$ and $j$ are   
adjacent on the grid.\ In this case,\ and
for $i<j$,\  we take the 
$J_{ij}$'s as  independent 
gaussian variables,\ with zero average  and
  variance $J^2 = 1$.\ This last choice fixes the (dimensionless) 
energy and temperature scales.\  
Neighbor configurations are by definition those
which differ in the   orientation of exactly one spin.
As an example, we show in figure~\mbox{1 a} the local density of states 
${\cal D}(E,L_{max})$
 for 25 realizations of the $J_{ij}$'s on the $5^3$ lattice.   
 We notice that ${\cal D}(E,L_{max})$ exhibits a rather simple behavior, 
 growing almost exponentially,
 with a systematic downward curvature. In a 
 semilogarithmic  plot the curvature is fully accounted for by a 
  parabola, which has 
 a  very small second order term. The density of states is for convenience
normalized to one within the pocket. The raw data are indicated by 
plusses and the full lines are fits
of the logarithm of $\cal{D}$ to a parabola.  
The second order term obtained in the fit is of the  order of
$1/50$ of the linear term.
Similarly, the available volume in the pocket is also close to an  exponential
 function of the lid energy, if one disregards the jumps which
 occur  whenever a new `side pocket' is accessed as the barrier increases. 

\subsection{Ferromagnetic Ising model} \label{ferromagnetic}
The Hamiltonian of the ferromagnetic Ising model  
has the same form as Eq.~\ref{Ham}, except for the      
crucial fact that   the non-zero $J_{ij}$ are all identical  
and equal to one. Again, neighbor configurations differ
by  one  spin flip.  This model's
landscape can hardly be considered as complex: it has two global 
energy minima, i.e. the ground states where all the spins are aligned,
and no local energy minima.   The lid algorithm was applied to 
the problem  in order to see whether the known 
critical temperature could be predicted from the form of
the local density of states. The latter quantity is
    depicted in figure~\mbox{1 b} for a 
$7 \times 7$, an $8 \times 8$ and a $4 \times 4 \times 4$
lattice. Remarkably, there is in all cases an exponential growth. 
The energy scales characterizing it 
are $2.66$, $2.70$ and $4.66$, respectively, which 
 compare favorably with the true
critical temperatures:  $2.27$ in $2d$\cite{Ma85} and 
$\approx 4.51$ in $3d$\cite{Domb74}.
 
\subsection{Structural glass (2d)} \label{structural}
A  random network of `atoms' placed 
on a 2D square lattice\cite{Schon97,Schon98} 
  can serve as a
model for covalent glasses\cite{Zachariasen}.
  The energy of a configuration
of the network is the sum of two-body- and three-body-potential
terms. The former has a repulsive term for
 short distances
 $r < 2d$, where $d$ is the lattice spacing, 
 and an attractive term, which  reaches zero smoothly for $r > 3.2d$.
The three-body-potential prevents bond angles below $80^\circ$, and  
introduces a preferred bond angle, which is    $\approx 120^\circ$. Thus, 
the global minimum configuration
would be a network consisting of  hexagons.  
To avoid surface effects, periodic boundary conditions are applied.
Neighbor configurations are created by shifting the position 
of one atom by one lattice unit.  
Each bit of the  binary string encoding a configuration
represents the state of one lattice site. Its value is one
if the site is occupied, and zero otherwise.
 
The energy landscapes of the  networks
were  investigated 
 for a range of densities 
 and sizes of the 
simulation cell. We show in 
figure~\mbox{1 c} a
typical example of $27$ atoms on a $14\times14$ lattice. The figure depicts
the accessible phase space volume ${\cal V}(L)$ and the number of accessible
minima ${\cal M}(L)$, together with the local density of states ${\cal D}(E;L)$,
and density of minima ${\cal DM}(E;L)$,
in a pocket enclosing a deep local minimum of the energy landscape. We note that all these
quantities show an approximately exponential growth with the lid $L$ and the
energy $E$, respectively.

\subsection{Polymer glass (2d)} \label{polymers}
As a final example we mention  random polymers on a two-dimensional 
lattice.  The
energy function is  similar  to that of the network model,
 except for two features:  The polymers are  not allowed to break
up or grow in size, i.e.  $V(r) = \infty$ for $r > 3.2$, and the interaction 
between atoms   not occupying  
consecutive  positions along a polymer
is purely repulsive for $r < 1.6$,  
and zero otherwise. 

As  the monomers have fixed positions
along different polymers, the encoding 
of a configuration is  more involved:
each  lattice point has a fixed  number, 
 and   each monomer  is assigned the binary 
value of  the  position it occupies in a given configuration. 
 These short binary strings are  appended to one another, in the order 
of  the building units of  the polymer.  This procedure creates the
long binary string used to identify the whole  configuration. 
  In figure~\mbox{1 d}, we show as one specific
example, the phase space volume ${\cal V}(L)$, the number of accessible minima
${\cal M}(L)$, the local density of states ${\cal D}(E;L)$ and the density of
minima ${\cal DM}(E;L)$ for a system of  two
 polymers of length
$18$ on a $14\times14$ lattice. We observe that the pocket in the 
landscape for these (relatively long)
polymers also exhibits exponential growth in the above quantities as a
function of $L$ and  $E$, respectively. 

\section{Summary and Discussion}
In this paper we have presented a rather general 
 algorithm for the exhaustive exploration
 of subsets of states (pockets) in complex energy 
landscapes. Such an exhaustive exploration can 
be used to map out the low energy part of energy 
landscape, yielding information which e.g.
 helps understanding
the low-temperature relaxation behavior of the systems. 
 We have   analyzed the program's 
 parallel performance, and demonstrated
 its applicability to four different 
 physical applications,  
which   required  a relatively large computational
effort.

The lid approach may  
complement  more traditional Monte Carlo simulations as 
an exploratory tool:  it yields complete information
on energetic barriers, but  is, by construction, insensitive to 
entropic barriers, e.g. bottlenecks between various parts of the
landscape which should all be "accessible" during relaxation
at low temperatures if only energetic barriers were  relevant.

It is interesting to note
that the landscapes of the rather different applications
considered share a key feature: the fact that
the local density of states within typical pockets 
is approximately exponential.
The energy  
scale of this exponential identifies the temperature at
which the trap looses its thermal
 metastability\cite{Schon97b,Sibani93,Sibani98}. 
In several instances 
this  temperature is also close to an actual transition temperature.
It thus appears that  exponential local densities of state within
 traps may
be part of the mechanism behind e.g. the spin glass
and glass transitions\cite{Sibani98,Schon98}.

As computers get faster and memory less expensive, it is our hope
that the lid algorithm will help to uncover the
landscape structure of many different complex systems, leading to a better
understanding of e.g. the dynamics of relaxations and 
phase transitions for such
systems. 

\noindent {\em Acknowledgments} The authors would like to 
thank the Silicon Graphics Advanced Technology Centre in 
Cortaillod (Switzerland) for making the Origin2000 system 
available to the project and  Statens Naturvidenskabelige
Forskningr\aa d for providing part of the computer resources
and for travel grants. P.S. is indebted to Richard Frost 
of the San Diego Supercomputing Center  for good advice 
on parallel computing, and C. S. would  like to thank the
 DFG for funding via SFB408 and a Habilitation stipend.

 \begin{table}
\begin{tabular}{l|l|l|l|r} 
Processors  & Elapsed time   & Rel. speed-up    & Cum. speed-up     &  Amdahl time     \\
\hline   
1   & 848   &  1.00  &  1.00   &  (848)                             \\
 2 &       506  & 1.68     &      1.68          &     (504)         \\
 4 &       298  &  1.70    &       2.85    &            (333)         \\
8 &       169   &  1.76    &       5.02    &            (247)    \\
16  &  115    &  1.47   &   7.37 &   (204)
\end{tabular}
\begin{caption}
\\
  Origin2000 performance for the 3-d spin-glass; lid value $L = 0.038$. \label{1}
\end{caption}
\end{table}

 \begin{table}
\begin{tabular}{l||l|l|l|l|r}   
                  
Function  &\multicolumn{5}{c }{ \# Processors}  \\
  & 1 & 2  &   4   &    8   &  16\\
 \hline
$generate$  &  479.3  &   247.9& 130.9 &  69.5& 38.3\\
$get\_from\_mail$ & 263.7 &  132.9 & 68.5 & 39.5 & 39.5 \\
$init\_energy3D$  & 102.9 &   52.1 & 26.5 & 13.3 &  6.6 \\ 
$flip$ & 7.4 & 4.2 &  2.3 &  1.4 &  0.9 \\
$mp\_barrier\_nthreads$ & 0.0 &   53.7 & 51.7 & 37.4 & 28.1 \\
\hline
Cumulative time     & 853.3 &  490.2 & 279.9 & 161.1 & 113.4
 
\end{tabular}
\begin{caption}
\\
Origin2000 performance breakdown at the function level
for the 3-d spin-glass; lid value $L = 0.038$. \label{2}
\end{caption}

\end{table}

 \begin{table} 
\begin{tabular}{l|l|l|l|r} 
Processors  & Elapsed time   & Rel. speed-up    & cum. speed-up     &  Amdahl time     \\
\hline   

   1    &    12535   &        1.00   &         1.00   &       (12535) \\
   2    &     7288   &        1.72   &         1.72   &        (7287) \\
   4    &     3914   &        1.86   &         3.20   &        (4664) \\
   8    &     2218   &        1.76   &         5.65   &        (3352) \\
  16    &     1342    &       1.65   &         9.34   &        (2696)

\end{tabular}
\begin{caption}
\\
  Origin2000 performance for the 3-d spin-glass; lid value $L = 0.046$.  \label{3}
\end{caption}
\end{table}

 \begin{table}
\begin{tabular}{l||r|r |r |r|r}   
                  
 Function &        \multicolumn{5}{c }{ Processors}  \\
 
 & 1 & 2  &   4   &    8   &  16\\
 \hline
$generate$           &       5685 & 3006 & 1552 & 811 &  435 \\
$get\_from\_mail$       &      5708 & 3650 & 1590 & 811 &  520 \\
$init\_energy3D$        &     1085 &  544 &  272 & 139 &   70 \\
$flip$                 &       95 &   54 &   31 &  17 &   11 \\
$mp\_barrier\_nthreads$  &      0 &  443 &  436 & 374 &  237 \\
\hline
Cumulative time     &      12573 & 7697 &  3881 &  2152 & 1273
\end{tabular}
\begin{caption}
\\
Origin2000 performance breakdown at the function
level for the 3-d spin-glass; lid value $L = 0.046$. \label{4}
\end{caption}

\end{table}  

 \begin{table} 
\begin{tabular}{l|r|r |r |r|r}   
                  
 Function &        \multicolumn{5}{c }{ Processors}  \\
 
 & 1 & 2  &   4   &    8   &  16\\
 \hline
$generate$            &      1.00 & 1.89 & 3.66 & 7.01 & 13.07 \\
$get\_from\_mail$        &     1.00 & 1.56 & 3.59 & 7.04 & 10.98 \\
$init\_energy3D$        &     1.00 & 1.99 & 3.99 & 7.81 & 15.50\\
$flip$                 &     1.00 & 1.76 & 3.06 & 5.59 &  8.64

\end{tabular}
\begin{caption}
\\
Parallel speed-up values for the 3-d spin-glass; lid value $L = 0.046$. \label{5} 
\end{caption}
\end{table}

 \begin{table}
\begin{tabular}{l|l|l|l|r} 
Processors  & Elapsed time   & Rel. speed-up    & Cum. speed-up     &  Amdahl time     \\
\hline   

   1    &    14011   &        1.00   &         1.00   &       (14011) \\
   2    &    10486  &        1.34   &         1.34  &        (7664) \\
   4    &     4455   &        2.35   &         3.15   &        (4490) \\
   8    &     2025   &       2.20   &         6.92  &        (2903)  
  
\end{tabular}
\begin{caption}
\\
  Origin2000 performance for 2-d network model; lid value $L = 4.2 (eV/atom)$. \label{6} 
\end{caption}
\end{table} 

\begin{figure}[t]
\vspace{-6cm}
%\centerline{\psfig{figure=fig1.eps}}
 \centerline{\psfig{figure=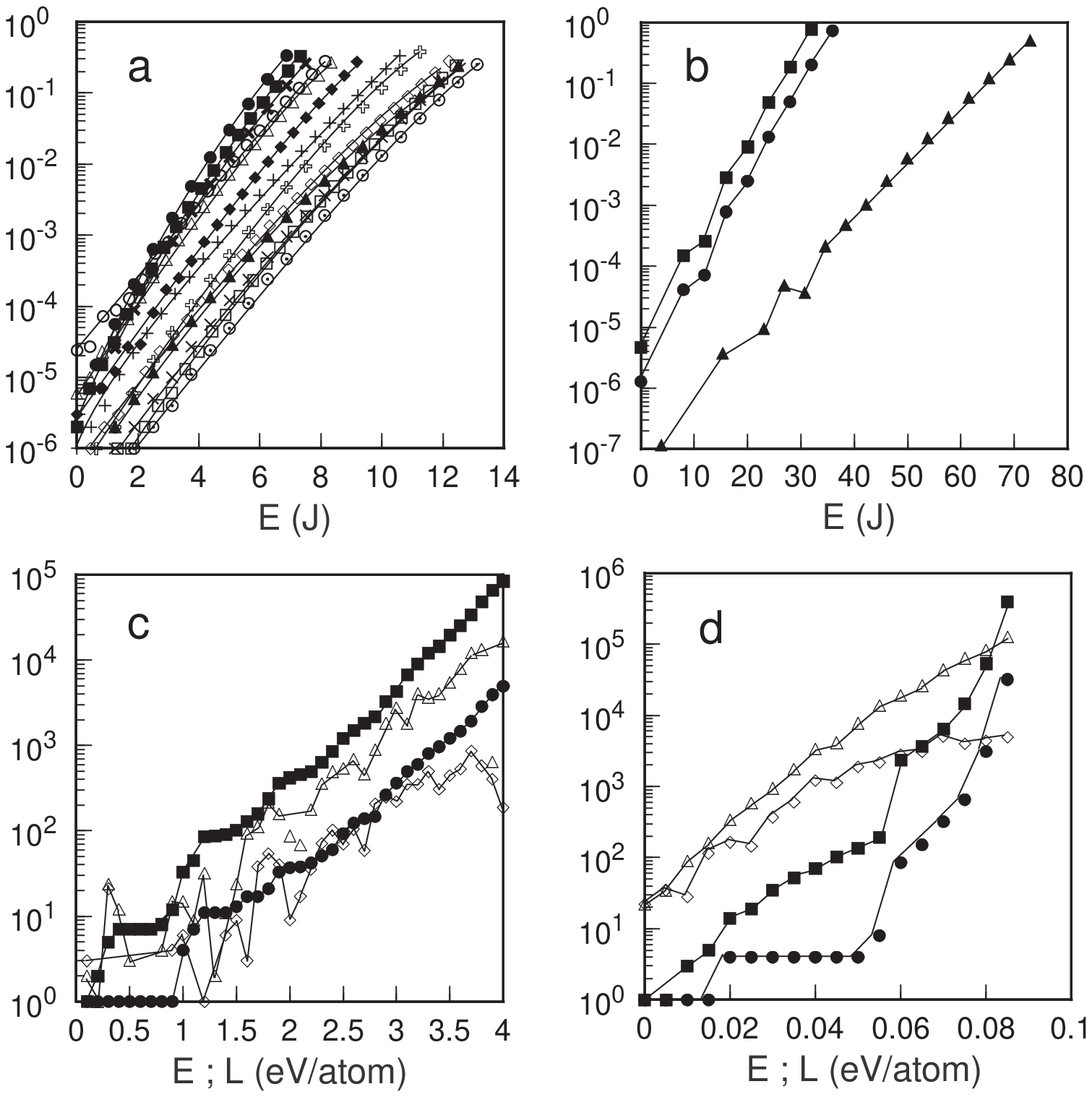,height=23.3cm,width=18cm}}
\vspace{-6.5cm} 
\begin{caption}  
\\   Plate  a) shows the  local densities of states 
${\cal D}(E,L_{max})$
 for 12 realizations of the $J_{ij}$'s of a spin-glass
 model on a  $5^3$ lattice. Plate b) shows the same quantity for
 the ferromagnetic Ising model. The circles are for 
 a $8\times 8$ lattice, the squares for   $7\times 7$ lattice and the
 triangles for a    $4\times 4\times 4$ lattice. In all cases 
 the data are divided by the total
 number of states found, which is seen to be of the order of 
  one million.  The abscissa is the total energy in units of 
 $J$.  For the spin glass $J$ is the standard deviation of the 
 distribution of the
 couplings  and the ferromagnet it is  the coupling constant itself.  
  The scale of the exponential growth parameter averaged
  over $25$ different realizations (13 data sets are omitted
  to avoid cluttering the graphics) is $T_c = 0.6$ for the spin glass model.
  In the ferromagnetic case we find  and
  $T_c \approx 2.7 $  in the two   dimensional lattices, 
  and $T_c \approx 4.7 $ in three dimensions.
  These figures are close to  the transition
  temperatures of the corresponding systems, which are $T_c \approx 0.84$
  for the spin glass,   $T_c = 2.27$ for the two-dimensional
  ferromagnet and  and $T_c \approx 4.51$ for the three dimensional one. 
  Plates c) and d) show data for the  2d network model of a glass 
  and for the polymer system, respectively. 
    The abscissa is either the energy $E$ or the 
  lid $L$, both in (eV/atom). The curves are : the available volume
  ${\cal V}(L)$ (squares) and the number of minima ${\cal M}(L)$ (circles)
  as a function of the lid; the local density of states  ${\cal D}(E,L_{max})$
  (triangles) and the local density of minima  ${\cal DM}(E,L_{max})$ (diamonds)
  as a function of the energy.
\end{caption}  
\end{figure}

\end{document}